\documentclass[rnote]{aa} 
\usepackage{graphicx}
\usepackage{natbib}
\usepackage{color}
\usepackage{txfonts}
\newcommand{\tw}{\color{black}}
\newcommand{\tv}{\color{black}}
\begin{document}
   \title{Testing non-linear force-free coronal magnetic field extrapolations with the
   Titov-D\'{e}moulin equilibrium.}

   \subtitle{}

   \author{T. Wiegelmann
          \inst{1}
          \and
          B. Inhester \inst{1}
          \and
          B. Kliem \inst{2}
          \and
          G. Valori \inst{2}
          \and
          T. Neukirch \inst{3}
          }

   \offprints{T. Wiegelmann}

   \institute{Max-Planck-Institut f\"ur Sonnensystemforschung,
Max-Planck-Strasse 2, 37191 Katlenburg-Lindau, Germany\\
              \email{wiegelmann@mps.mpg.de}
         \and
             Astrophysical Institute Potsdam,
             An der Sternwarte 16, 14482 Potsdam, Germany\\
             \and
       School of Mathematics and Statistics,
   University of St. Andrews,
   St. Andrews, KY16 9SS,
   United Kingdom \\
             }

   \date{A\&A, Vol. 453, 737-741 (2006)}


  \abstract
   {
   As the coronal magnetic field can usually not be measured directly, it has to
   be extrapolated from photospheric measurements into the corona.
   }
   {We test the quality of a non-linear force-free coronal magnetic field
   extrapolation code with the help of a known {{\tv analytical}} solution.}
  {
  The non-linear force-free equations are {{\tv numerically}} solved with the help of an optimization
   principle. The method minimizes an integral over the force-free and solenoidal condition.
   As boundary condition we use either the magnetic field components on all six sides of the
   computational box in Case I or only on the bottom boundary in Case II. We check the quality
   of the reconstruction by computing how well force-freeness and divergence-freeness are fulfilled
   and by comparing the {{\tv numerical}} solution with the {{\tv analytical}} solution. The comparison is
   done with magnetic field line plots and several quantitative measures, like the vector correlation,
   Cauchy Schwarz, normalized vector error, mean vector error and magnetic energy.
   }
   {For Case I the reconstructed magnetic field shows good agreement with the original magnetic
   field topology, whereas in Case II there are considerable deviations from the exact solution.
   This is corroborated by the quantitative measures, which are significantly better for Case I.}
   {
   Despite the strong nonlinearity of the considered force-free equilibrium,
   the optimization method of extrapolation is able to reconstruct it;
   however, the quality of reconstruction depends significantly on the  consistency
   of the input data, which is given only if the known solution  is provided also at
   the lateral and top boundaries, and on the presence  or absence of flux concentrations
   near the boundaries of the magnetogram.
   }

 \authorrunning{Wiegelmann et al.}
 \titlerunning{Testing magnetic field extrapolations}
   \maketitle
%

\section{Introduction}
Several methods have been proposed to compute the non-linear force-free
coronal magnetic field in active regions from measurements of
the photospheric magnetic field vector, e.g. the direct upward
integration method \citep{wu:etal90}, the Grad-Rubin method
{\tw \citep[e.g.][]{sakurai81,amari:etal99,regnier:etal02,bleybel:etal02,wheatland04,amari:etal06,inhester:etal06}}, the
Green's function like methods \citep{yan:etal00}, the stress and relax
method \citep{roumeliotis96,valori:etal05} and the optimization
method \citep{wheatland:etal00,wiegelmann04,wiegelmann:etal06}.

A standard test for force-free extrapolation methods is the application of the codes
to known analytical or numerical nonlinear force-free equilibria. Due to the general
mathematical difficulty of the problem, only very few such solutions are known. The
comparison of numerical nonlinear force-free extrapolation codes with the class of
nonlinear force-free equilibria found by \citet{low:etal90} (LL from here on) has in
the past ten years emerged as a certain standard test. The LL equilibria are a class
of axisymmetric equilibria which are separable in spherical coordinates. They are
self-similar in the radial coordinate, and the polar angle dependence is determined
from a non-linear eigenvalue equation. The symmetry is broken by cutting out a
rectangular chunk of the solution by using a Cartesian coordinate system which is
shifted  and rotated with respect to the original coordinate system in which the LL
equilibria are calculated. The parameters of the LL solutions and the parameters of
the new Cartesian coordinate system allow for a large number of different situations
which can be used for tests.

Nevertheless, it is highly desirable to also use tests different from LL. A first
step has been made by \citet{valori:etal05}, who used an equilibrium taken from a
numerical investigation of a twisted flux rope. {\tv Valori and Kliem (2005, private
communication) have further tested their nonlinear force-free extrapolation code
with configurations containing a flux rope by considering the approximate analytical
force-free equilibrium established by \cite[][henceforth TD]{titov:etal99}.}
In a recent review talk on nonlinear force-free extrapolation,
T. Neukirch also strongly recommended the use of other equilibria, in
particular the TD-equilibrium,  to test the quality of non-linear
force-free extrapolation codes. The TD equilibrium  has a much more
concentrated current density than LL, and the magnetogram is more structured.
Also, the final equilibrium has a different topology than the corresponding potential field, which
is not true for LL.

The aim of this research note is to study the performance
 the optimization method in reconstructing the TD equilibrium.
 The optimization code has so far only been tested with the
LL solution with rather small resolution ($40 \times 40 \times 20$)  in
\citet{wheatland:etal00}, on $80 \times 80 \times 40$ grids in
\citet{wiegelmann:etal03} and on a $64^3$ grid as part of a comparison of six
different extrapolation methods in \citet{schrijver:etal06}. The optimization code in
the implementation of \citet{wiegelmann04} was the fastest-converging and
best-performing model of the six tested extrapolation codes. Here we test this
optimization code  with the TD equilibrium using a high numerical resolution ($150 \times
250 \times 100$).

\section{The TD equilibrium}

Like the LL solutions, the TD equilibrium is an axisymmetric equilibrium and the symmetry is broken by choosing the boundaries appropriately.
The TD equilibrium has originally been constructed as a model of an active region
containing a current- carrying flux tube. The model has later been used in a number of investigations
of the initiation of CME eruptions \citep[e.g.,][]{roussev:etal03,toeroek:etal04,kliem:etal04,toeroek:kliem05, williams:etal05}.

The axis of symmetry of the TD equilibrium is placed at a distance $d$
beneath the lower boundary (the photosphere). A line current of strength $I_0$ runs along
the line of symmetry, which creates a potential magnetic field with circular field lines.
\citet{titov:etal99} then added a toroidal nonlinear force-free current around the axis of
symmetry with  minor radius $a$ and major radius $R$ and total current $I$, assuming $a \ll R$
(see also Fig. 2 in \cite{titov:etal99}).
Similar to a tokomak, the ring current would not be in equilibrium with the potential field
created by the line current. Therefore, two magnetic monopoles (strength $q$) of opposite
polarity are placed on the symmetry axis at a distance $L$ from each other, with the
force-free ring current half-way between them. The monopoles create a potential {\tv poloidal} magnetic
{\tv field} which has field lines lying above the force-free ring current and thus holding it down.
In this way a stable equilibrium situation can be achieved for certain parameter combinations
(unstable cases have been investigated in the studies of CME initiation mentioned above).

\begin{table}
\caption{Parameter set of the TD equilibrium}
\label{parameters}
\begin{tabular}{lll}
   $R$ &=110 Mm      = $2.2$   &  (major torus radius)  \\
   $a$ & = 35 Mm      = $0.7$   &  (minor torus radius) \\
   $d$ & = 50 Mm      = $1.0$   &  (depth of torus center) \\
   $L$ & =100 Mm      = $2.0$   &  (monopole distance) \\
   $q$ & =100 T~Mm$^2$    &  (magnetic charge) \\
   $I_0$ & =-13 TA               & (line current) \\
   $ I$ &  = $3.391$ TA   $   $ &  (ring current) \\
   $h_\mathrm{apex}$  &        = $1.2$   &  (apex height) \\
   $y_\mathrm{foot}$  &        = $1.960$ &  (footpoint position) \\
   $B_\mathrm{apex}$  &        = $1.0$   &  (norm. apex field strength) \\
   $<\Phi>$  & = $-1.41 \pi$       & (average twist) \\
\end{tabular}
\end{table}

{\tv For the present investigation we use the TD equilibrium (given by
their equations 16--22 and 31) with the set of parameters given in
Table~\ref{parameters}. This yields a stable equilibrium because the
twist of the flux rope is clearly sub-critical with respect to the kink
instability \citep{toeroek:etal04} and the monopoles are placed
sufficiently distant \citep{titov:etal99}.}
In Table \ref{parameters} we have also {\tv included} the derived parameters of the apex height of
the ring current above the photosphere ($=R-d$),
the footpoint position of the ring current ($=\sqrt{R^2-d^2}$),
the average field line twist in the nonlinear force-free ring current (averaged over the {\tv toroidal}
cross section),
{\tv and the normalization of the field and of the various length parameters}.

The TD equilibrium has a number of properties which we mention here explicitly because
they are relevant for magnetic field extrapolation methods. The artificial magnetogram
of the TD equilibrium basically has a large-scale bipolar structure (due to the line current),
but this structure is modified by the monopoles and the ring current to an almost quadrupolar structure.
Because the line current extends from $-\infty$ to $\infty$, the magnetic field on the photosphere
does not drop to zero with larger distance from the nonlinear force-free region
{\tv in the direction of the line current}. This is different
from the LL solutions, for which the photospheric magnetic field can be made very small
{\tv at the edges of the magnetogram} if the computational box is chosen large enough.
On the other hand, the coronal current density (and therefore the region with non-vanishing $\alpha$ on
the photospheric boundary) is strongly concentrated for the TD equilibrium (to the region of the ring current).
In the LL case $\alpha$ is distributed over extended regions of the lower boundary. Finally, as already mentioned above,
the TD equilibrium magnetic field has a different topology than the corresponding potential magnetic field calculated from
the $B_z$ component of the magnetic field on the lower boundary. This is not the case for the standard LL cases usually used
for testing nonlinear force-free extrapolation codes.

\section{Method}
\label{method}
Force-free coronal magnetic fields have to obey the equations
\begin{eqnarray}
(\nabla \times {\bf B })\times{\bf B} & = & {\bf 0},  \label{forcefree}\\
\nabla\cdot{\bf B}    & = &         0      \label{solenoidal-ff}.
\end{eqnarray}
We solve these equations with the help of an optimization principle, as proposed
by \citet{wheatland:etal00} and generalized by \citet{wiegelmann:etal03a}.
We define the functional
\begin{equation}
L=\int_{V}  w(x,y,z) \, \left[B^{-2} \, |(\nabla \times {\bf B}) \times {\bf
B}|^2 +|\nabla \cdot {\bf B}|^2\right] \, d^3x
\label{defL1},
\end{equation}
where $w(x,y,z)$ is a weighting function.  It is obvious that (for $w>0$) the
force-free Eqs. (\ref{forcefree}-\ref{solenoidal-ff}) are fulfilled when $L$ equals
zero.  We compute the magnetic field in a box with $nx=150$, $ny=250$ and $nz=100$
points.  The numerical method works as follows.  As an initial configuration we
compute a potential magnetic field in the whole box with the help of a Green's
function method as described in \citet{aly89}. The next step is slightly different
for the two cases.
\begin{itemize}
\item Case I: We impose the
exact TD magnetic field vector on all six boundaries of the computational box. The weighting function is $w=1$ in the
entire box.
\item Case II: We impose the
exact TD magnetic field vector only on the bottom boundary (photosphere).
The boundary conditions on the lateral and top boundaries
of the computational box are given by the initial potential field.
The weighting function is
$w=1$ in the center $118 \times 218 \times 84$ region and drops to $0$ with a cosine profile
in a $16$-pixel boundary layer towards the lateral and top boundaries of the computational box.
\end{itemize}
In both cases we iterate for the magnetic field inside the
computational box by minimizing Eq. (\ref{defL1}).
The program is written in C and has been parallelized with OpenMP. The computations
have been done on 8 Procs. The details of the current implementation of our code
are described in \citet{wiegelmann04}.

\begin{table*}
\caption{Figures of merit for the comparison of our reconstruction with the original
TD equilibrium. See Sec. \ref{sec4} for details.
The diagnostic was done on the center $118 \times 218 \times 84$ grid and
a boundary layer of $16$ points towards the lateral and top boundaries of
the computational box has been omitted.}
\label{table1}
\centering
\begin{tabular}{|l|l|l|l|l|l|l|l|l|l|l|l|r|r|}     
\hline Model & $L_{\rm inner}$&$L_{1 \, {\rm inner}}$&$L_{2 \, {\rm inner}}$& {\tw
$\parallel \nabla \cdot {\bf B} \parallel_{\infty}$}&{\tw $\parallel {\bf j } \times
{\bf B} \parallel_{\infty} $}&
$C_{\rm vec}$&$C_{\rm CS}$&$E_{N}^{'}$&$E_{M}^{'}$&$\epsilon$&$\epsilon_P$&It. Steps& Comp. Time \\
\hline
T \& D &$0.015$&$0.013$&$0.002$&$3.13$&$5.54$&$1$&$1$&$1$&$1$&$1$&$2.36$&-&-\\
Potential &$0.26$&$5.5 \, 10^{-8}$&$0.26$&$11.89$&$0.076$&$0.82$&$0.82$&$0.39$&$0.35$&$0.42$&$1$&-&22 min\\
Case I &$0.0016$&$0.0012$&$0.0004$&$1.92$&$2.36$&$0.9996$&$0.9998$&$0.99$&$0.99$&$0.999$&$2.36$&22180&6h 37 min\\
Case II &$0.37$&$0.34$&$0.02$&$2.88$&$10.48$&$0.96$&$0.94$&$0.66$&$0.58$&$0.66$&$1.56$&5330&1h 35 min\\
\hline
\end{tabular}
\end{table*}

   \begin{figure*}
   \centering
\mbox{\includegraphics[clip,bb=40 20 384 375,height=7.0cm,width=8cm]{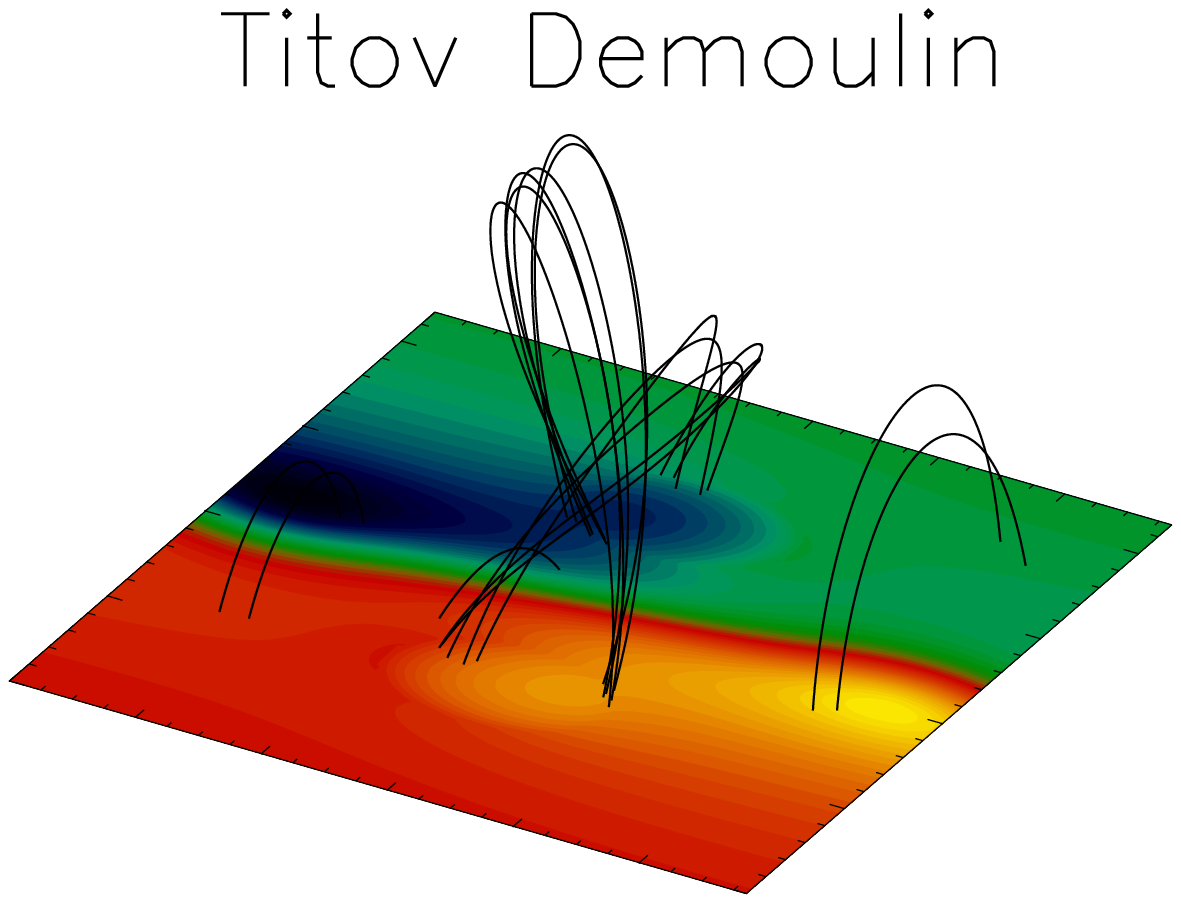}
   \includegraphics[clip,bb=40 20 384 375,height=7.0cm,width=8cm]{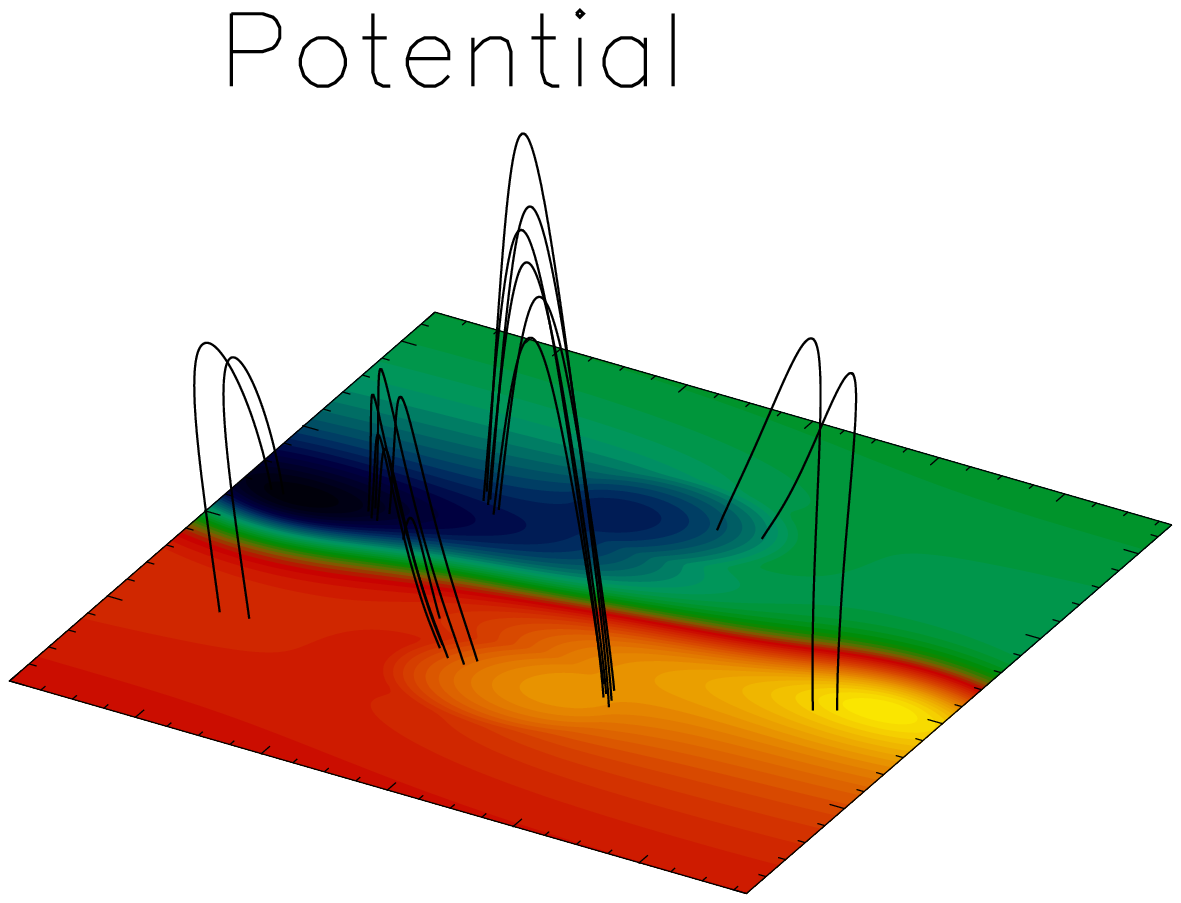}}
\mbox{\includegraphics[clip,bb=40 20 384 375,height=7.0cm,width=8cm]{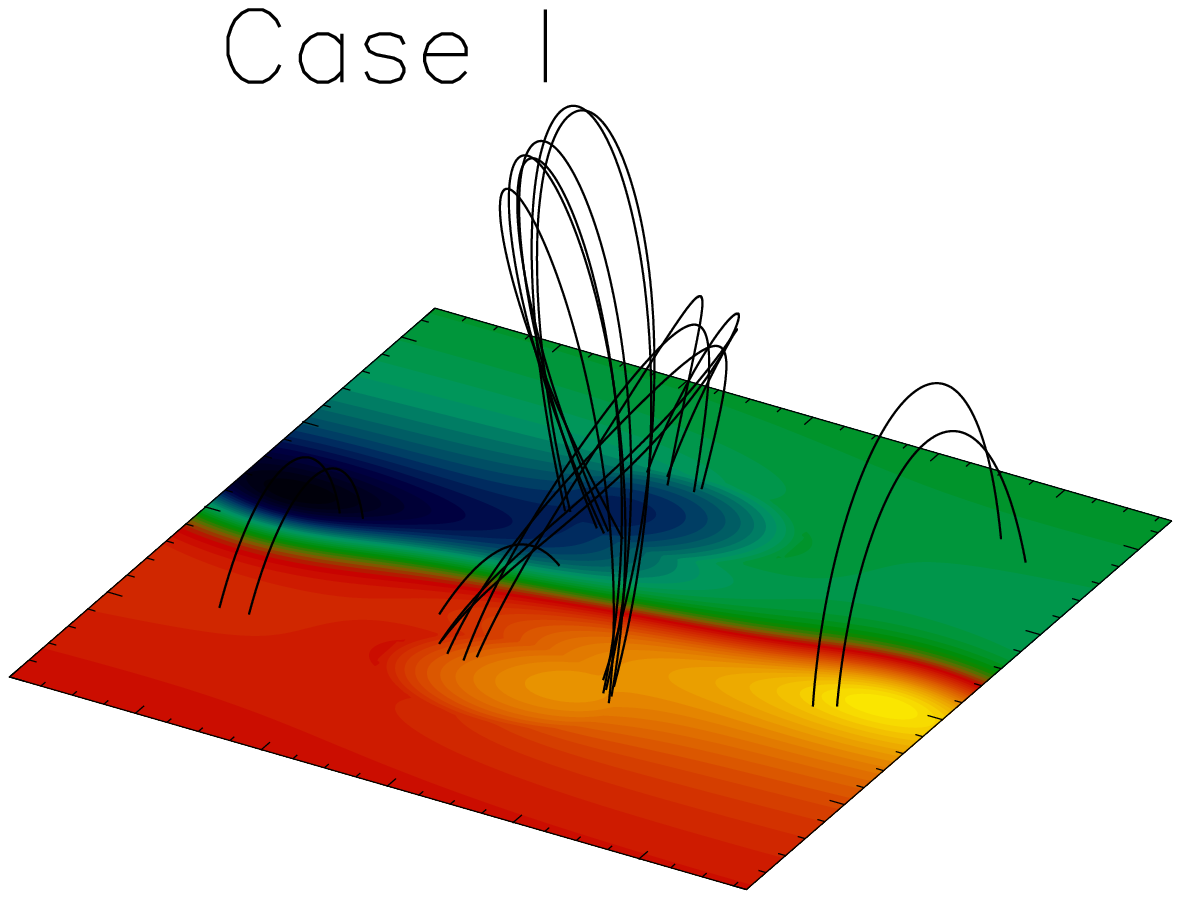}
   \includegraphics[clip,bb=40 20 384 375,height=7.0cm,width=8cm]{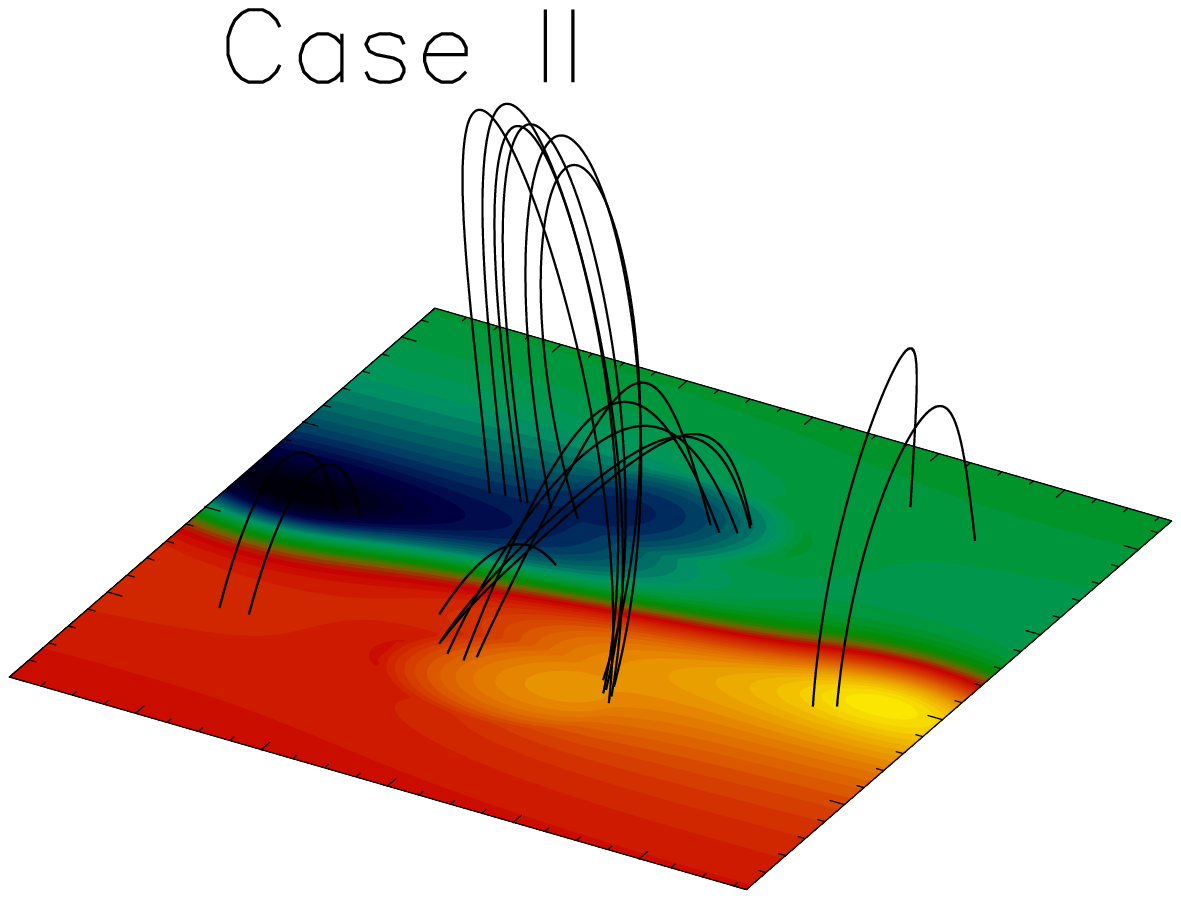}}
   \caption{We compare the magnetic field structure of the original Titov Demoulin
   equilibrium (Titov Demoulin) with a potential field extrapolation (Potential),
   a non-linear force-free extrapolation with all six boundary surfaces prescribed (Case I) and
   a non-linear force-free extrapolation which uses only the photospheric magnetic field data
   (Case II).
   The panels show the center $118 \times 218 \times 84$ region of the original
   $150 \times 250 \times 100$.
   The colour coding shows the magnetic field strength on the photosphere. We used
   the same start points for the magnetic field line computation in the positive (yellow)
   magnetic field region on the photosphere in all panels.
    }
              \label{fig1}%
    \end{figure*}

\section{Figures of merit}
\label{sec4} To assess the quality of our reconstruction we introduce several
figures of merit in Table \ref{table1}. The first column of the table gives the name
of model, column two contains the integral $L=L_1+L_2$ defined in Eq. \ref{defL1},
the third column $L_1$ shows how good the force-free condition (first term of the
integral in Eq. \ref{defL1}) is fulfilled  and the forth column $L_2$ corresponds to
the solenoidal condition (second term of the integral in Eq. \ref{defL1}). $L$,
$L_1$ and $L_2$ are calculated with the magnetic field normalized by the average
magnetic field strength on the photosphere and the length scale normalized by the
average box length.
 {\tw
 We emphasize, however, that we restrict the integration domain to the inner part of the
 computational box, excluding the boundary layer in which $w(x,y,z) \ne 1$.
 Thus strictly speaking the diagnostic quantities differ from the integrals
 given in Eq. \ref{defL1}. We therefore call them $L_{\rm inner}$,
 $L_{1 \, {\rm inner}}$ and $L_{ 2 {\rm inner}}$.}
Columns five to {\tw eleven} contain quantitative measures on how good the
reconstructed solution agrees with the original Titov \& Demoulin equilibrium. These
figures of merit have been introduced by \citet{schrijver:etal06} to compare the
results of six different extrapolation codes (including the code used here) with the
Low \& Lou solution. {\tw In a coordinated study \citet{amari:etal06} used the same
figures of merit to compare the reconstruction of two Grad-Rubin codes also with the
Low \& Lou solution.} These figures quantify the agreement between vector fields
${\bf B}$ (Titov \& Demoulin equilibrium) and ${\bf b}$ (reconstructed fields). We
use exactly the same definitions as given in Sect.~4 of \citet{schrijver:etal06}, in
order to allow also an comparison of how well our code reconstructs the Low \& Lou
solution (see Tables~I and II in \citet{schrijver:etal06}) and the Titov \& Demoulin
solution (Table \ref{table1} here).

{\tw
 Column five contains the $L_{\infty}$ norm of the divergence of the magnetic
 field
\begin{equation}
\parallel \nabla \cdot {\bf B} \parallel_{\infty}=
\sup_{{\bf x} \in V} |\nabla \cdot {\bf B}|
\end{equation}
Column six contains the $L_{\infty}$ norm of the Lorentz force of the magnetic
 field
\begin{equation}
\parallel {\bf j } \times {\bf B} \parallel_{\infty}=
\sup_{{\bf x} \in V} |{\bf j } \times {\bf B} |
\end{equation}

}
 Column seven contains the vector correlation
\begin{equation}
C_{\rm vec}=  \sum_i {\bf B_i} \cdot {\bf b_i}/
\left( \sum_i |{\bf B_i}|^2 \sum_i |{\bf b_i}|^2 \right)^{1/2},
\end{equation}
column eight the Cauchy-Schwarz inequality
\begin{equation}
C_{\rm CS} = \frac{1}{N} \sum_i \frac{{\bf B_i} \cdot {\bf b_i}}
{|{\bf B_i}||{\bf b_i}|},
\end{equation}
where $N$ is the number of vectors in the field.
The normalized vector error is defined as
\begin{equation}
E_{\rm N} = \sum_i |{\bf b_i}-{\bf B_i}|/ \sum_i |{\bf B_i}|,
\end{equation}
and the mean vector error as
\begin{equation}
E_{\rm M} = \frac{1}{N} \frac{\sum_i |{\bf b_i}-{\bf B_i}|}{\sum_i |{\bf B_i}|}.
\end{equation}
Columns nine and ten contain $E^{'}_N = 1-E_N$ and $E^{'}_M =1-E_M$, respectively.
In column eleven we show the total magnetic energy of the reconstructed field
normalized with the energy of the input field
\begin{equation}
\epsilon = \frac{\sum_i |{\bf b_i}|^2}{\sum_i |{\bf B_i}|^2}.
\end{equation}
The two vector fields agree perfectly if the figures of merit ($C_{\rm vec},C_{\rm
CS},E^{'}_N,E^{'}_M,\epsilon$) are unity. In the twelves column we list $\epsilon_P$
which is the magnetic energy normalized by the energy of the corresponding potential
field. Column thirteen contains the number of iteration steps until convergence and
column fourtenn the computing time on 8 processors.

 %
\section{Results}
 In Fig. \ref{fig1} we compare the original TD equilibrium
 with a potential field extrapolation (Potential), a non-linear force-free
 extrapolation with all six boundaries prescribed (Case I) and a non-linear
 force free extrapolation where only the bottom boundary has been prescribed
 (Case II). The potential field does obviously not agree with the original.
 Case I shows a very good agreement with the original.
 Case II reconstructs the magnetic field approximately. Deviations
 of the magnetic field lines between Case II and the original are visible, in particular
 for field lines emerging in the nonlinear force-free region of the bottom boundary.

 The visible inspection of Fig. \ref{fig1} is supported by the figures
 of merit in Tab. \ref{table1}. The figures of merit
 (columns 7-11) are for case I better
 than $1 \%$ for all figures, mostly even better than $0.1 \%$.
 The integral force and divergence free conditions (columns 1-3)
 are even one order of magnitude better than the discretization
 error of the exact solution. The $L_{\infty}$ norms (columns 4 and 5)
 are of the same order of magnitude as the the discretization error.

  The reason is that the optimization
 code minimizes $L$ with respect to the numerical grid. For case II,
 on the contrary, $L$ is one order of magnitude higher than the discretization
 error of the exact solution.
 The reason is that for case II an inconsistency exists between the bottom  boundary
 and the other boundaries, where the initial potential field is  kept. Correspondingly,
 the figures of merit are less good than in case I.

The lateral and upper boundaries try to feed information about the original
potential field into the solution, and because the potential field has a different
topology, the flux rope of the TD equilibrium is not correctly reproduced in
Case~II.
 Fig. 2
(online material) shows that the field lines calculated from start points lying on
symmetric circles at the footpoints of the flux rope do outline a single flux rope
only in Case~I, but outline two flux ropes that separate with height in Case~II.


%
\section{Conclusions}
 We used the TD equilibrium to test a non-linear force-free
 optimization code. In case I we prescribed all six boundaries of
 the computational box and in case II only the bottom boundary.
 For case I we get an almost perfect agreement with the original,
 and for case II also a reasonable agreement.
 Case I is the real performance test of the code, because only if the
 correct boundary conditions on all boundaries are prescribed we have exactly the
 same physical problem and can expect to find the exact solution.
 For the reconstruction of magnetic fields from
 observed vector magnetograms the results of Case II
 are more important, however, because for real active regions the
 lateral boundaries are unknown.
 Let us also remark that the figures of merit here are comparable with
 the figures of merit in the LL case (Table I  rows
 b) in \citet{schrijver:etal06}). For cases II  the results for
 TD are
 slightly worse for $L, L_1, C_{\rm vec},E^{'}_N  $ and  somewhat better  for
 $L_2, C_{\rm CS}, E^{'}_M $ compared with the LL case. The reconstructed
 magnetic energy in the TD case discussed here is not as accurate
 (only for case II) as for the LL case. One has to consider,
 however, that the magnetic energy of the corresponding LL field
 \citep[case II in][]{schrijver:etal06} is only $1.10$ times the potential
 field, where the TD equilibrium has a magnetic energy which
 is $2.36$ higher than the potential field. The magnetic field strength
 on the photosphere for the LL field becomes very low close to the
 lateral  boundaries, which is not the case for TD. Despite
 these difficulties, our code reconstructed the magnetic field topology
 approximately correct, even if the lateral boundaries are unknown.
 Current vectormagnetograms often have a limited field
 of view and consequently also significant magnetic flux close to the
 lateral boundaries. A comparison of direct measurements of chromospheric
 magnetic loops
 by \citet{solanki:etal03} with a non-linear force-free extrapolation from
 a photospheric vectormagnetogram with a limited field of view in
 \citet{wiegelmann:etal05} showed
 a reasonable agreement, except for loops close to boundaries of the
 magnetogram.

\begin{acknowledgements}
The work of T.\ Wiegelmann was supported by DLR-grant 50 OC 0501 and a
 British Council-DAAD
 grant at the University of St. Andrews.
 T.\ Wiegelmann acknowledges the warm hospitality of the St. Andrews Solar Theory
 Group during two research visits.
 B.\ Kliem and G.\ Valori were supported by grants DFG MA
 1376/16-2 and DFG HO 1424/9-1, respectively.
We acknowledge useful discussions during two workshops on non-linear force-free
field extrapolation methods organized by Karel Schrijver in Palo Alto 2004 and 2005.
We thank the referee, Tahar Amari, for useful remarks to improve this paper.
\end{acknowledgements}

\bibliographystyle{aa}

\begin{figure*}
\setlength{\unitlength}{1.0cm}
\begin{picture}(14,21)
\put(0,12){\includegraphics[height=6.0cm,width=8cm]{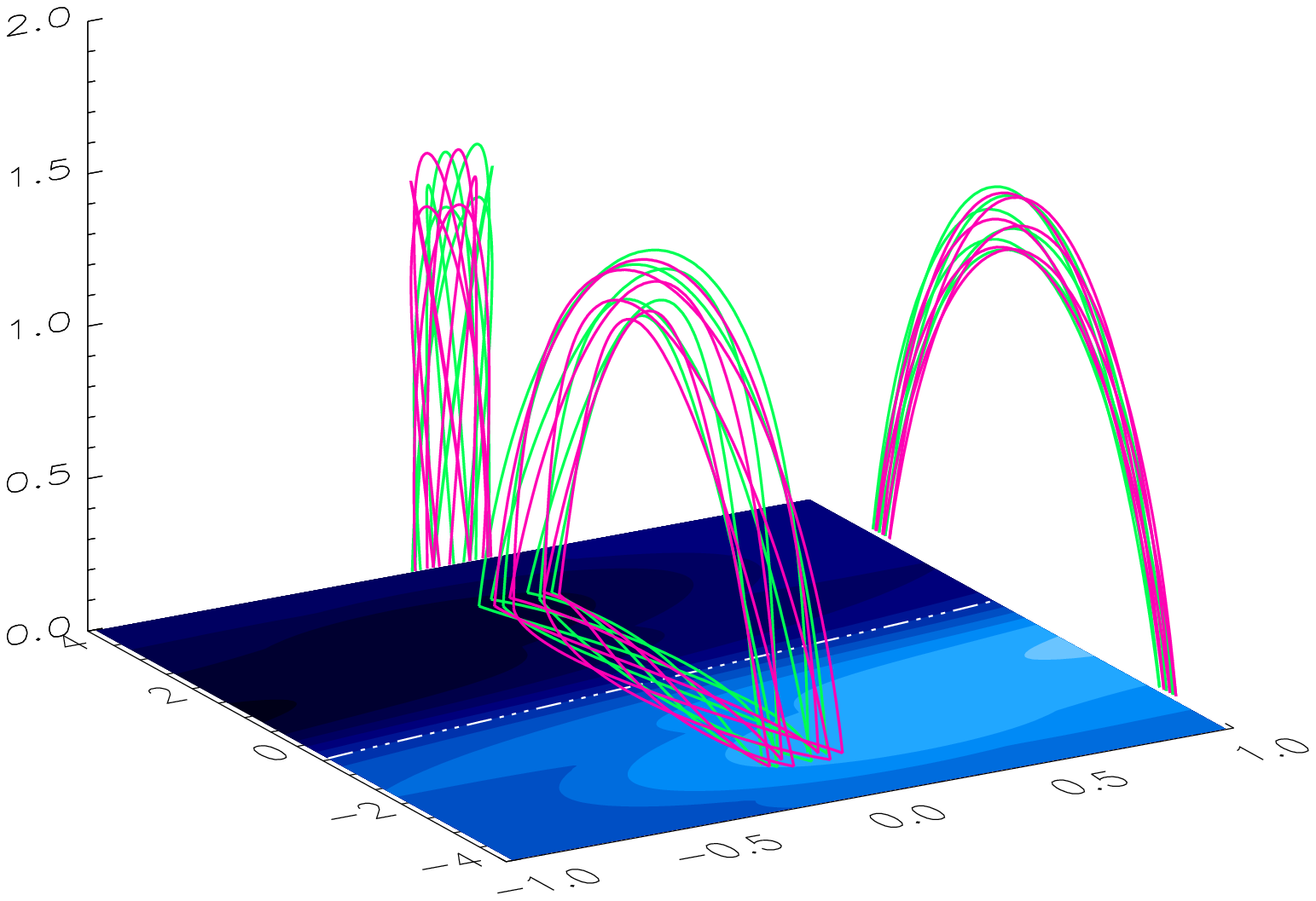}}
\put(0,6){\includegraphics[height=6.0cm,width=8cm]{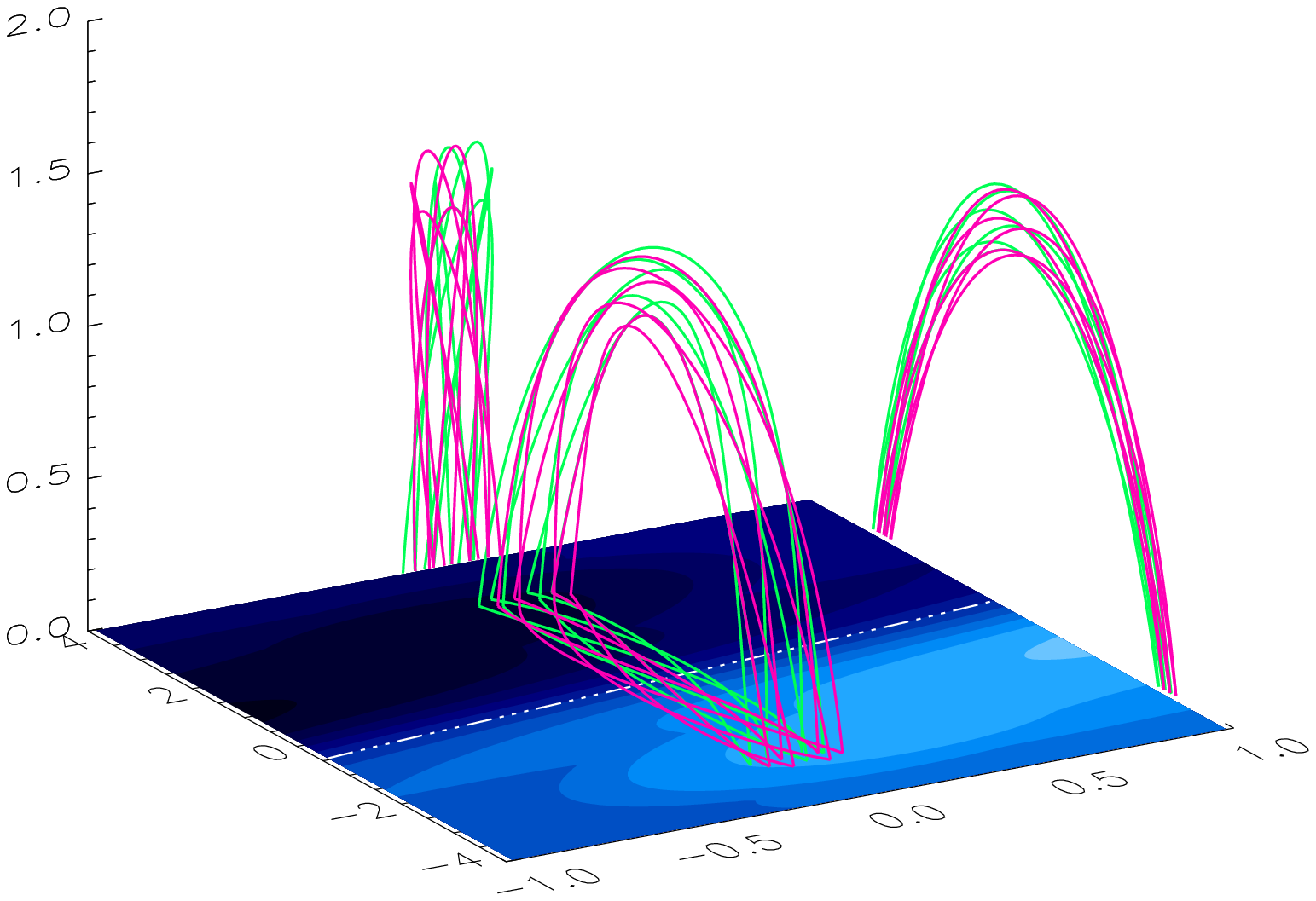}}
\put(0,0){\includegraphics[height=6.0cm,width=8cm]{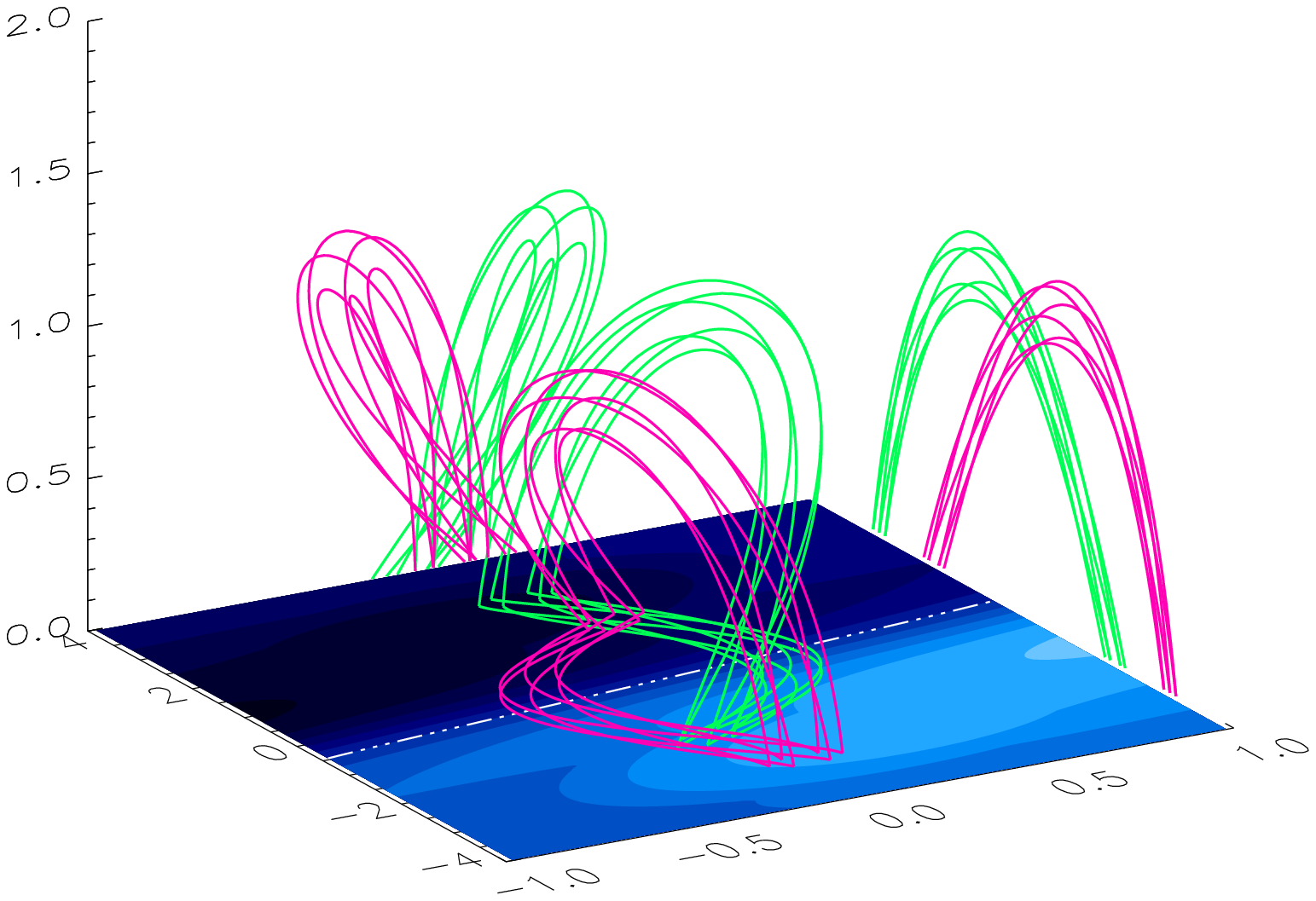}}
\put(2,17){\parbox{5.5cm}{{\bf \LARGE Titov Demoulin}}}
\put(2,11){\parbox{5.5cm}{{\bf \LARGE Case I}}}
\put(2,5){\parbox{5.5cm}{{\bf \LARGE Case II}}}
\put(0,20){\parbox{12.0cm}{{\bf \Huge Online material}}}
\put(10,10){\parbox{5.0cm}{\caption{The TD equilibrium has a current ring surrounded by potential
field. Being force-free this means that there is a  semi-circular flux rope
in correspondence of the current ring. Here we check how good the extrapolated final topology
reproduces that. We plot two sets of 6 field lines, each set defining a flux tube.
The field lines start from 6 equally spaced points on two circles of radius
$0.1$, each centered at one of the intersections of the current ring axis with
the magnetogram $( x=z=0, y=\pm y_\mathrm{foot} )$.
The aim of such a plot is to visualize the flux rope that sustains the current
ring in TD.
In the original TD field the two flux tubes overlap completely: one ends where the
other starts.
Case I confirms that the extrapolation using the
information on all six boundaries recovers the flux rope to an excellent
degree of accuracy.
However, when only one boundary is used for the extrapolation the two flux
tubes do not coincide (Case II).}}}
\end{picture}

\label{online_fig2}
\end{figure*}

\end{document}